\documentclass[twocolumn,aps,prl,showpacs,superscriptaddress]{revtex4-1}

\usepackage[latin1]{inputenc}
\usepackage{latexsym}
\usepackage{amsmath}
\usepackage{amssymb}

\def\endproof{\vrule height6pt width6pt depth0pt}

\begin{document}


\title{State-independent quantum contextuality for continuous variables}


\author{\'{A}ngel R. Plastino}
 \email{arplastino@ugr.es}
 \affiliation{Instituto Carlos I de F\'{\i}sica Te\'orica y
 Computacional and Departamento de F\'{\i}sica At\'omica,
 Molecular y Nuclear, Universidad de Granada, E-18071 Granada,
 Spain}


\author{Ad\'{a}n Cabello}
 \email{adan@us.es}
 \affiliation{Departamento de F\'{\i}sica Aplicada II,
 Universidad de Sevilla, E-41012 Sevilla, Spain}


\date{\today}



\begin{abstract}
Recent experiments have shown that nature violates
noncontextual inequalities regardless of the state of the
physical system. So far, all these inequalities involve
measurements of dichotomic observables. We show that
state-independent quantum contextuality can also be observed in
the correlations between measurements of observables with
genuinely continuous spectra, highlighting the universal
character of the effect.
\end{abstract}



\maketitle


{\em Introduction. }Recent experiments \cite{KZGKGCBR09,
BKSSCRH09, ARBC09, LHGSZLG09, MRCL10}, following the proposal
in \cite{Cabello08}, have shown that nature cannot be described
by noncontextual theories and that this impossibility can be
detected by a state-independent violation of an inequality. The
motivation behind these experiments comes from the observation
made by Kochen and Specker \cite{Specker60, KS67}, and Bell
\cite{Bell66} that contextuality is a necessary property in any
attempt to complete quantum mechanics (QM) with additional
variables. In a similar way, nonlocality is a necessary
property in any attempt to complete QM \cite{Bell64}. However,
while nonlocality is only needed to explain the quantum
predictions when the physical system is prepared in an
entangled state, contextuality is needed to explain the quantum
predictions regardless of which state the system is in. The
recent developments reported in \cite{KZGKGCBR09, BKSSCRH09,
ARBC09, LHGSZLG09, MRCL10, Cabello08, Cabello09} are helping to
overcome the obstacles for the experimental study of quantum
contextuality that have been pointed out in the literature
\cite{Meyer99, Kent99, CK00, Cabello02, BK04} (see
\cite{CKCLKZGR10} for a detailed discussion).

So far, all state-independent violations of noncontextual
inequalities \cite{KZGKGCBR09, ARBC09, MRCL10, Cabello08,
BBCP09, Cabello09, Cabello10} invoke dichotomic observables.
Moreover, the original proofs of the impossibility of
noncontextual alternatives to QM are only valid for observables
with discrete spectra \cite{KS67, Bell66}. The case of
continuous variables like position or momentum is of
fundamental importance since ``all measurements of quantum
mechanical systems could be made to reduce eventually to
position and time measurements'' \cite{FH65} or, according to
Bell, ``in physics the only observations we must consider are
position observations'' \cite{Bell82}. There is an extensive
literature on Bell inequalities for local hidden-variable
theories with continuous variables \cite{BW99, CPHZ02, CFRD07}.
Additional motivations behind these researches are: (a) to
extend the range of quantum states violating the inequalities
and achieve the maximal violation \cite{CPHZ02}, and (b) to
avoid the need of dichotomic binning of the results to get a
violation \cite{CFRD07}. On the other hand, the extension of
quantum information to continuous variables has attracted great
interest, since it has important technological implications
\cite{LB99, BP03, BV05, WTSTD09}.

Therefore, a fundamental question is whether there is a
state-independent violation of a noncontextual inequality using
only continuous variables. In this article we derive a simple
noncontextual inequality for continuous variables such that,
according to QM, there is a universal set of observables for
which: (a) any state maximally violates the inequality, and (b)
the violation does not require any binning of the results.


{\em Inequality. }Consider $18$ observables, $A'$, $A''$, $B'$,
$B''$, $C'$, $C''$, $a'$, $a''$, $b'$, $b''$, $c'$, $c''$,
$\alpha'$, $\alpha''$, $\beta'$, $\beta''$, $\gamma'$, and
$\gamma''$ which take any possible value between $-1$ and $1$:
\begin{subequations}
\begin{align}
&-1 \le A' \le 1,\ldots,\\
&-1 \le \gamma'' \le 1.
\end{align}
\end{subequations}
In addition, these values are assumed to satisfy the following
restrictions:
\begin{subequations}
\label{normalization}
\begin{align}
&|A'+i A''|=1,\ldots,\\
&|\gamma'+i\gamma''|=1,
\end{align}
\end{subequations}
where $i$ is the imaginary constant and $|x|$ denotes the
modulus of the complex number $x$. For convenience, hereafter
we use the following notation:
\begin{subequations}
\begin{align}
&A=A'+iA'',\ldots,\\
&\gamma=\gamma'+i\gamma'',
\end{align}
\end{subequations}
and we consider mean values like $\langle A B C \rangle =
\langle (A'+iA'') (B'+iB'') (C'+iC'') \rangle$, where $A'$,
$A''$, $B'$, $B''$, $C'$, and $C''$ are mutually compatible
observables. To experimentally obtain $\langle A B C \rangle$,
one has to sequentially measure the six observables on the same
individual system and then compute the complex number
$(A'+iA'') (B'+iB'') (C'+iC'')$. Then, one must repeat the
experiment many times on identically prepared copies and take
the average of the real part and the average of the imaginary
part.

{\em Lemma: }Any theory in which all these $18$ observables
have predetermined noncontextual outcomes (i.e., independent of
which compatible observables are jointly measured) must satisfy
the following inequality:
\begin{equation}
 |\langle ABC \rangle
 +\langle abc \rangle
 +\langle \alpha \beta \gamma \rangle
 +\langle A a \alpha \rangle
 +\langle B b \beta \rangle
 -\langle C c \gamma \rangle|
 \le 3 \sqrt{3},
 \label{inequality}
\end{equation}
where the observables inside each mean value are mutually
compatible.

{\em Proof: }To obtain the upper bound of inequality
\eqref{inequality}, let us first find the maximum possible
value of
\begin{equation}
\begin{split}
 |S| &= |A B C + a b c + \alpha \beta \gamma + A a \alpha + B b \beta
 - C c \gamma|,\\
 &= | A (B C + a \alpha) + b (a c + B \beta)+ \gamma (\alpha \beta - C c)|,
\end{split}
 \label{efe}
\end{equation}
where $A$, $B$, $C$, $a$, $b$, $c$, $\alpha$, $\beta$, and
$\gamma$ are nine arbitrary complex numbers of modulus $1$.
Then,
\begin{equation}
|S| \le |B C + a \alpha| + |a c + B \beta |
+ |\alpha \beta - C c|,
\label{modefe}
\end{equation}
since $|A|=|b|=|\gamma|=1$.

To find an upper bound for the right-hand side of
\eqref{modefe}, we introduce the phases $\phi_1 $ and $\phi_2$,
defined as
\begin{subequations}
\label{phases}
\begin{align}
&\frac{BC}{a \alpha} = e^{i \phi_1}, \\
&\frac{ac}{B \beta} = e^{i \phi_2}, \\
&\frac{Cc}{\alpha \beta} = e^{i (\phi_1 + \phi_2)}.
\end{align}
\end{subequations}
From \eqref{phases}, it follows that,
\begin{subequations}
\label{trigon}
\begin{align}
&|B C + a \alpha|^2 = 4 \cos^2 \frac{\phi_1}{2}, \\
&|a c + B \beta |^2 = 4 \cos^2 \frac{\phi_2}{2}, \\
&|\alpha \beta - C c|^2 = 4 \sin^2\left[\frac{1}{2}\left(\phi_1+\phi_2\right)\right].
\end{align}
\end{subequations}
From Eqs. \eqref{trigon} we see that finding the maximum of the
right-hand side of \eqref{modefe} is tantamount to finding the
maximum of
\begin{equation}
 2 \left\{\left|\cos \frac{\phi_1}{2}\right| + \left|\cos \frac{\phi_2}{2}\right| + \left|\sin\left[\frac{1}{2}(\phi_1 + \phi_2)\right]\right|\right\}.
 \label{trigonextrem}
\end{equation}
It can be easily seen that this maximum is $3 \sqrt{3} \approx
5.19$ (for instance, it occurs when $\phi_1=\phi_2=\pi/3$).
Therefore,
\begin{equation}
|S| \le 3 \sqrt{3}.
\end{equation}

Finally, if one repeats the experiment many times on
identically prepared copies of the system, then one can use
that
\begin{equation}
|\langle S \rangle| \le \langle |S| \rangle
\end{equation}
and obtains inequality \eqref{inequality}.\hfill\endproof

The upper bound of \eqref{inequality} can be reached, for
instance, for
\begin{subequations}
\begin{align}
&A'=C''=b'=c''=\gamma'=\frac{\sqrt{3}}{2},\\
&A''=-C'=b''=-c'=-\gamma''=-\frac{1}{2},\\
&B'=a'=\alpha'=\beta'=1,\\
&B''=a''=\alpha''=\beta''=0.
\end{align}
\end{subequations}

{\em Quantum violation. }Let us consider a quantum-mechanical
system consisting of a particle moving in a two-dimensional
space. This system has continuous position and momentum
observables, ${\bf x} = (x_1,x_2)$ and ${\bf p} = (p_1,p_2)$,
that comply with the standard canonical commutation relations,
\begin{subequations}
\label{xp}
\begin{align}
&[x_i,x_j]=0,\\
&[p_i,p_j]=0,\\
&[x_i,p_j]= i \hbar \delta_{ij}.
\end{align}
\end{subequations}

Now consider the $18$ observables described in QM by the
following Hermitian operators:
\begin{subequations}
\label{observables}
\begin{align}
 &A'=\cos\left(\frac{p_0}{\hbar} x_1 \right),\;\;\;\;\;\;\;\;\;\;\;\;\;\;
 A''=\sin\left(\frac{p_0}{\hbar} x_1 \right),\\
 &B'=\cos\left(\frac{\pi}{p_0} p_2\right),\;\;\;\;\;\;\;\;\;\;\;\;\;
 B''=\sin\left(\frac{\pi}{p_0} p_2\right),\\
 &C'=\cos\left(\frac{p_0}{\hbar} x_1 + \frac{\pi}{p_0} p_2 \right),\;
 C''=-\sin\left(\frac{p_0}{\hbar} x_1 + \frac{\pi}{p_0} p_2 \right),\\
 &a'=\cos\left(\frac{p_0}{\hbar} x_2 \right),\;\;\;\;\;\;\;\;\;\;\;\;\;\;\;\;
 a''=-\sin\left(\frac{p_0}{\hbar} x_2 \right),\\
 &b'=\cos\left(\frac{\pi}{p_0} p_1 \right),\;\;\;\;\;\;\;\;\;\;\;\;\;\;\;\;
 b''=\sin\left(\frac{\pi}{p_0} p_1\right),\\
 &c'=\cos\left(\frac{p_0}{\hbar} x_2 - \frac{\pi}{p_0} p_1 \right),\;\;\;
 c''=\sin\left(\frac{p_0}{\hbar} x_2 - \frac{\pi}{p_0} p_1 \right),\\
 &\alpha'=\cos\left[\frac{p_0}{\hbar}(x_2-x_1)\right],\;\;\;\;\;
 \alpha''=\sin\left[\frac{p_0}{\hbar}(x_2-x_1)\right],\\
 &\beta'=\cos\left[\frac{\pi}{p_0}(p_1+p_2)\right],\;\;\;\;\;
 \beta''=-\sin\left[\frac{\pi}{p_0}(p_1+p_2)\right],\\
 &\gamma'=\cos\left[\frac{p_0}{\hbar}(x_1-x_2) + \frac{\pi}{p_0}(p_1+p_2) \right],\nonumber \\
 &\gamma''=\sin\left[\frac{p_0}{\hbar}(x_1-x_2) + \frac{\pi}{p_0}(p_1+p_2) \right],
\end{align}
\end{subequations}
where $p_0$ is a constant with dimensions of momentum. These
$18$ observables are examples of {\em modular variables} which
have played a distinguished role in the interpretation of the
Aharonov-Bohm and related effects \cite{APP69}, the study of
the Greenberger-Horne-Zeilinger proof for continuous variables
\cite{MP01}, and in dynamic quantum nonlocality
\cite{Popescu10}.

The $18$ observables \eqref{observables} comprise six sets of
six mutually compatible observables. For instance, the six
observables in $\langle A B C \rangle = \langle (A'+iA'')
(B'+iB'') (C'+iC'') \rangle$ are mutually compatible; therefore
the corresponding operators commute. This mutual commutativity
is evident for the observables appearing in the products $A B
C$, $a b c$, $\alpha \beta \gamma$, $A a \alpha$, and $B b
\beta$. Indeed, in each of these cases the six observables are
of the form $\cos \Theta$, $\sin \Theta$, $\cos \Xi$, $\sin
\Xi$, $\cos [-(\Theta+\Xi)]$, and $\sin [-(\Theta+\Xi)]$, where
$[\Theta,\Xi]=0$. Moreover, these five products are of the form
\begin{equation}
 e^{i \Theta} e^{i \Xi} e^{-i(\Theta+\Xi)}=\openone.
 \label{PM1}
\end{equation}
Therefore, according to QM,
\begin{equation}
 \langle A B C \rangle
 =\langle a b c \rangle
 =\langle \alpha \beta \gamma \rangle
 =\langle A a \alpha \rangle
 =\langle B b \beta \rangle=1.
 \label{obstruction1}
\end{equation}
Interestingly, observables $C'$, $C''$, $c'$, $c''$, $\gamma'$,
and $\gamma''$ are also compatible. This property can be
derived from the following argument. Standard canonical
commutation relations \eqref{xp} imply Weyl's canonical
commutation relations (see, e.g., \cite{EMS04}),
\begin{equation}
\label{Weyl}
\begin{split}
 \exp\left(-\frac{i}{\hbar} r x_i\right) \exp\left(-\frac{i}{\hbar} t p_i\right)
 = & \exp\left(-\frac{i}{\hbar}r t\right) \exp\left(-\frac{i}{\hbar} t p_i\right)
 \\
 & \times \exp\left(-\frac{i}{\hbar} r x_i\right).
\end{split}
\end{equation}
An important particular instance of these relations is obtained
when
\begin{equation}
 r t = \pm 2 \pi \hbar.
 \label{dospi}
\end{equation}
In this case, \eqref{Weyl} reduces to
\begin{equation}
 \exp\left(-\frac{i}{\hbar} r x_i\right)\exp\left(-\frac{i}{\hbar} t p_i\right)
 =\exp\left(-\frac{i}{\hbar} t p_i\right)\exp\left(-\frac{i}{\hbar} r x_i\right).
 \label{Weylplus}
\end{equation}
From \eqref{dospi} and \eqref{Weylplus}, it follows that
\begin{subequations}
\label{trigxp}
\begin{align}
 &\left[\cos\left(\frac{r}{\hbar} x_i \right),\cos\left(\frac{t}{\hbar} p_i \right)\right]=0,\\
 &\left[\sin\left(\frac{r}{\hbar} x_i \right),\sin\left(\frac{t}{\hbar} p_i \right)\right]=0.
\end{align}
\end{subequations}
By the same token, the relation
\begin{equation}
\left[
\frac{1}{\sqrt{2}}
\left(
x_1 + \frac{\hbar \pi}{p_0^2} p_2
\right),
 \frac{1}{\sqrt{2}}
\left(
- \frac{p_0^2}{\hbar \pi} x_2
+ p_1
\right)
\right] = i \hbar
\end{equation}
implies that $[C', c']=0$. The commutativity between the rest
of the operators in the set $C'$, $C''$, $c'$, $c''$,
$\gamma'$, and $\gamma''$ can be deduced in a similar way. The
significance for fundamental issues in quantum mechanics of the
fact that appropriate trigonometric functions of two
observables may commute even if these observables do not was
first pointed out in \cite{APP69}. The common eigenbasis
associated with each of the sets of commuting operators
considered here are described in the Appendix.

Now let us calculate
\begin{equation}
\label{Ccgamma}
 \begin{split}
 C c \gamma = &
 \exp\left[-i\left(\frac{p_0}{\hbar} x_1 + \frac{\pi}{p_0} p_2 \right)\right]
 \exp\left[i\left(\frac{p_0}{\hbar} x_2 - \frac{\pi}{p_0} p_1 \right)\right] \\
 & \times \exp\left\{i\left[\frac{p_0}{\hbar}(x_1-x_2) + \frac{\pi}{p_0}(p_1+p_2) \right]\right\}.
 \end{split}
\end{equation}
Since $[x_1,p_2]=0$, $[x_2,p_1]=0$, $[x_1-x_2,p_1+p_2]=0$,
$[x_1,x_2]=0$, and $[p_1,p_2]=0$, the exponentials on the
right-hand side of \eqref{Ccgamma} can be factorized as
\begin{equation}
\label{Ccgamma2}
 \begin{split}
 &\exp\left(-i\frac{p_0}{\hbar} x_1\right) \exp\left(-i\frac{\pi}{p_0} p_2\right)
 \exp\left(i\frac{p_0}{\hbar} x_2\right) \exp\left(-i\frac{\pi}{p_0} p_1\right) \\
 &\times \exp\left(i\frac{p_0}{\hbar} x_1\right) \exp\left(-i\frac{p_0}{\hbar} x_2\right)
 \exp\left(i\frac{\pi}{p_0} p_1\right) \exp\left(i\frac{\pi}{p_0} p_2\right).
 \end{split}
\end{equation}
Now we repeatedly apply another special instance of Weyl's
relations \eqref{Weyl},
\begin{equation}
 \exp\left(-\frac{i}{\hbar} r x_i\right) \exp\left(-\frac{i}{\hbar} t p_i\right)
 =-\exp\left(-\frac{i}{\hbar} t p_i\right) \exp\left(-\frac{i}{\hbar} r x_i\right),
 \label{Weylminus}
\end{equation}
corresponding to
\begin{equation}
 r t = \pm \pi \hbar.
 \label{pi}
\end{equation}
Using three times \eqref{pi} and \eqref{Weylminus} in
\eqref{Ccgamma2}, we obtain
\begin{equation}
 C c \gamma = -\openone.
 \label{PM2}
\end{equation}
Consequently, according to QM,
\begin{equation}
 \langle C c \gamma \rangle =-1.
 \label{obstruction2}
\end{equation}
Therefore, from \eqref{obstruction1} and \eqref{obstruction2},
the quantum-mechanical prediction for $|S|$ is
\begin{equation}
 |\langle S_{\rm QM} \rangle|=6,
\end{equation}
which violates the upper bound of inequality
\eqref{inequality}, $|\langle S \rangle|\le 3 \sqrt{3} \approx
5.19$. This violation is {\em maximal} and is the same for {\em
any} state of the system, even for mixed states regardless
their degree of mixture. The state independency of the
violation is particularly interesting for continuous-variable
systems where it is generally difficult to prepare specific
states.

Equations \eqref{PM1} and \eqref{PM2} indicate that the nine
unitary operators $A=A'+iA'',\ldots,\gamma=\gamma'+i\gamma''$
formally behave like those of the celebrated Peres-Mermin
square of two-qubit operators \cite{Peres90, Mermin90}. A
similar observation was made by Clifton \cite{Clifton00}.


{\em Conclusions. }No fundamental difficulty seems to exist to
observe a state-independent violation of noncontextual
inequalities with continuous variables. For any quantum system
admitting two continuous position observables, $x_1$ and $x_2$,
and the corresponding canonically conjugate momenta, $p_1$ and
$p_2$, we have shown that there exists a set of universal
observables with continuous spectra which can experimentally
reveal state-independent quantum contextuality. The observables
$x_1$, $x_2$, $p_1$, and $p_2$ could represent, for instance,
the position and momentum of a particle moving in a
two-dimensional space, or the positions and momenta of two
particles, each of them moving in a one-dimensional space
\cite{EPR35}, or the quadrature amplitudes of two modes of the
electromagnetic field \cite{GDR98}. The required measurements,
although discussed long ago in the literature, are probably
hard to implement in actual experiments using specific physical
systems, and this issue demands further research. But the
important point is that, according to QM, there is no
fundamental obstacle to carry out these measurements and
observe the effect.


\begin{acknowledgments}
A.R.P. acknowledges support from MCI Project No.~FIS2008-02380
and Junta de Andaluc\'{\i}a Excellence Project
No.~P06-FQM-02445. A.C. acknowledges support from MCI Project
No.~FIS2008-05596 and Junta de Andaluc\'{\i}a Excellence
Project No.~P06-FQM-02243.
\end{acknowledgments}


\section{Appendix}


In this Appendix we provide the common eigenbasis associated
with each of the sets of commuting operators considered in the
paper. Following standard, self-explanatory notation, we denote
by $|x_{1,2}\rangle$ and $|p_{1,2}\rangle$ the eigenstates of
the position and momentum operators $x_{1,2}$ and $p_{1,2}$,
respectively. These eigenstates are normalized in the standard
way, $\langle x_1|x_1^{\prime}\rangle =
\delta(x_1-x_1^{\prime})$, etc. It is plain that the common
eigenbasis corresponding to the sets $\{A,B,C\}$, $\{a,b,c\}$,
$\{A,a,\alpha\}$, and $\{B,b,\beta\}$ consist of states
 (again using self-explanatory notation)
$|x_1\rangle|p_2\rangle$, $|p_1\rangle|x_2\rangle$,
$|x_1\rangle|x_2\rangle$, and $|p_1\rangle|p_2\rangle$,
respectively. The common eigenbasis of the compatible operators
$\{\alpha, \beta, \gamma\}$ is constituted by states of the
form $|x_{-}\rangle|p_{+}\rangle$, where $|x_{-}\rangle$ and
$|p_{+}\rangle$ stand for eigenstates of the operators $x_{-} =
x_2 - x_1$ and $p_{+} = p_1 + p_2$, respectively. If we
interpret $x_1$ and $x_2$ as the coordinates of two particles
of equal mass, then $x_{-}=x_2-x_1$ and $x_{+} =
\frac{1}{2}(x_1 + x_2)$ represent the relative and
center-of-mass coordinates, and $p_{-}= \frac{1}{2}(p_2-p_1)$
and $p_{+} = p_1 + p_2$ the concomitant canonically conjugate
momenta.

Let us now consider the common eigenbasis of the set of
operators $\{C,c,\gamma\}$. In order to construct this
eigenbasis, it is convenient to first introduce the observables
\begin{subequations}
 \begin{align}
 &V_1 = x_1 + \frac{\pi \hbar}{p_0^ 2}\, p_2, \,\,\,\,\,\,
 W_1= -\frac{p_0^ 2}{2\pi\hbar}\, x_2 + \frac{p_1}{2}, \\
 &V_2 = x_2 + \frac{\pi \hbar}{p_0^ 2}\, p_1, \,\,\,\,\,\,
 W_2= -\frac{p_0^ 2}{2\pi\hbar}\, x_1 + \frac{p_2}{2}.
 \end{align}
\end{subequations}
These observables satisfy the commutation relations,
$[V_j,W_k]=i\hbar \delta_{jk}$, $[V_j,V_k]=[W_j,W_k]=0$, for
$j,k=1,2$. That is, the observables $V_{1,2}$ and $W_{1,2}$
comply with the same commutation relations as $x_{1,2}$ and
$p_{1,2}$. In the standard $(x_1,x_2)$ coordinate
representation, the common eigenstates $|v_1,v_2\rangle$ of
$V_1$ and $V_2$ are given by the wave function
\begin{equation}
 \label{autoves}
 \langle x_1, x_2 | v_1, v_2 \rangle = \frac{1}{2\pi}
 \exp \left[
 i (v_2 x_1 + v_1 x_2) -
 \frac{i p_0^ 2}{\pi \hbar^2} x_1 x_2
 \right],
\end{equation}
complying with the normalization condition $\langle v_1, v_2 |
v_1^{\prime}, v_2^{\prime} \rangle = \delta(v_1-v_1^{\prime})
\delta(v_2-v_2^{\prime})$. The eigenvalues corresponding to the
observables $V_{1,2}$ associated with eigenstate
\eqref{autoves} are $\frac{\pi \hbar^2}{p_0^ 2}v_{1,2}$. The
eigenstates $|v_1,v_2\rangle$ constitute a (continuous) basis
for the Hilbert space describing the system under
consideration.

Now we can express the operators $C$ and $c$ in terms of $V_1$
and $W_1$,
\begin{subequations}
 \begin{align}
 &C = \exp \left(-\frac{i}{\hbar} p_0 V_1\right),\\
 &c = \exp \left[-\frac{i}{\hbar} \left(\frac{2\pi \hbar}{p_0} \right)W_1
 \right]. \label{CV}
 \end{align}
\end{subequations}
It follows from the commutation relation verified by the
observables $V_j$ and $W_j$ and from \eqref{CV} that the
operator $c$ represents a displacement in the ``direction''
$v_1$,
\begin{equation}
 \label{displa}
 c |v_1,v_2\rangle = \Big|v_1 +
 \frac{2 p_0 }{\hbar}, v_2\Big\rangle.
\end{equation}
The commutation relations verified by the operators $V_j$ and
$W_j$ [and, in particular, relation \eqref{displa}] imply that
the common eigenbasis of the commuting operators $C$, $c$, and
$V_2$ is given by the states
\begin{equation}
 \label{commonbas}
 |\kappa, \varepsilon, v_2 \rangle =
 \frac{1}{\sqrt{2\pi}} \sum_{n=-\infty}^{+ \infty}
 e^{i \kappa n} \Big|\varepsilon + \frac{2 p_0 }{\hbar} n, v_2\Big\rangle,
\end{equation}
where $\kappa \in [0,2\pi)$, $\varepsilon \in \left[0, \frac{2
p_0 }{\hbar}\right)$, and $-\infty < v_1 < +\infty$. The states
$\Big|\varepsilon + \frac{2 p_0 }{\hbar} n, v_2\Big\rangle$ in
\eqref{commonbas} are common eigenstates of $V_1$ and $V_2$,
given by wave functions of the form \eqref{autoves}. The
eigenvalues of the eigenstate \eqref{commonbas} associated with
the operators $C$, $c$, and $V_2$ are, respectively,
\begin{equation}
 \exp \left(-\frac{i\pi \hbar \varepsilon}{p_0}\right), \,\,\,\,\,\,\,\,
 \exp(- i \kappa),\,\,\,\,\,\,\,\, \frac{\pi \hbar^2 v_2}{p_0^2}.
\end{equation}
The eigenstates \eqref{commonbas} are normalized as
\begin{equation}
 \langle \kappa, \varepsilon, v_2 | \kappa^{\prime},
 \varepsilon^{\prime}, v_2^{\prime} \rangle =
 \delta(\kappa - \kappa^{\prime})
 \delta(\varepsilon - \varepsilon^{\prime})
 \delta(v_2 - v_2^{\prime}).
\end{equation}

Since the operators $C$, $c$, and $\gamma$ satisfy the relation
$\gamma C c = - \openone$ [which can be derived in the same way
as \eqref{PM2}], it follows that the state $|\kappa,
\varepsilon, v_2 \rangle$ is also an eigenstate of the operator
$\gamma$, with eigenvalue
\begin{equation}
 - \exp \left[i \left(\kappa + \frac{\pi \hbar \varepsilon}{p_0}
 \right) \right].
\end{equation}
We have shown that the states $|\kappa, \varepsilon, v_2
\rangle$ constitute a common eigenbasis of the operators $C$,
$c$, and $\gamma$. It is easy to verify that those states are
also eigenstates of the six compatible observables,
$C^{\prime}$, $C^{\prime \prime}$, $c^{\prime}$, $c^{\prime
\prime}$, $\gamma^{\prime}$, and $\gamma^{\prime \prime}$, with
eigenvalues $\cos \left( \frac{\pi \hbar \varepsilon}{p_0}
\right)$, $-\sin \left( \frac{\pi \hbar \varepsilon}{p_0}
\right)$, $\cos (\kappa)$, $-\sin(\kappa)$, $-\cos \left(
\kappa + \frac{\pi \hbar \varepsilon}{p_0} \right)$, and $-\sin
\left( \kappa + \frac{\pi \hbar \varepsilon}{p_0} \right)$,
respectively. Finally, it can be easily seen that the
orthonormal states \eqref{commonbas} constitute a basis of the
Hilbert space of the system under study. To see that, it is
enough to verify that the states $|v_1,v_2\rangle$ given by the
wave functions \eqref{autoves} (which clearly constitute a
basis) can be expressed as linear combinations of the states
\eqref{commonbas}. Indeed, if we set $\varepsilon = v_1 -
\frac{2p_0}{\hbar} m$, with $m$ equal to the integer part of
$\frac{\hbar v_1}{2 p_0}$,
\begin{equation}
 \label{integerpart}
 m = {\rm int} \left(\frac{\hbar v_1}{2 p_0} \right),
\end{equation}
we have,
\begin{equation}
 \label{basisatlast}
 |v_1,v_2\rangle = \int_{0}^{2\pi}
 \frac{d\kappa}{2\pi} \,\, e^{-i \kappa m} \,
 |\kappa,\varepsilon, v_2 \rangle.
\end{equation}



\end{document}